\documentclass{revtex4}
\usepackage{amsmath}
\usepackage{amssymb}
\usepackage{graphicx}

\begin{document}

\title{Bianchi type I model with two interacting scalar fields}

\author{Vladimir Folomeev}
\email{vfolomeev@mail.ru} \affiliation{Institute of Physics of NAS
KR, 265 a, Chui str., Bishkek, 720071,  Kyrgyz Republic}

\begin{abstract}
The flat anisotropic model of the early Universe is considered. Two
interacting scalar fields with special form of potential energy are a source of matter fields.
Analytic solutions for
inflationary and scalaron stages are found. It is shown that the solutions tend asymptotically to
isotropic Friedmann model.

%\bigskip
%\noindent Keywords:
\end{abstract}

%\pacs{ }

\maketitle

\section{Introduction}
Investigation of models of the early Universe has a long history (see e.g. Refs.~\cite{Linde,Lidd}). One of the basic
direction is consideration of classical scalar fields. For such type of models the necessary
inflationary evolution can be ensured by a corresponding choice of energy-momentum tensor
which is equivalent to the hydrodynamic equation of state $p=-\varepsilon$. In this case the
exponential expansion of the Universe is realized, and its size increase from the Planck value up to
macroscopic one. The subsequent redistribution of the field energy between kinetic and potential
parts leads to fast oscillations of the field (the so-called scalaron stage~\cite{Star}) and to realization
of the effective equation of state $p\ll \varepsilon$. Such fast oscillations with the Planck energy
density provide mass creation of elementary particles with subsequent filling of the Universe by
hot plasma. At the corresponding choice of initial parameters, such a simple model solves the principal
problems cosmology (causality, homogeneity, flatness, e.t.c).

In this Letter we offer the model of the early Universe using the special form of the potential energy of scalar field:
\begin{equation}
\label{pot2}
V(\phi,\chi)=\frac{\lambda_1}{4}(\phi^2-m_1^2)^2+\frac{\lambda_2}{4}(\chi^2-m_2^2)^2+\phi^2 \chi^2-V_0.
\end{equation}
Here $\phi, \chi$ are two real scalar fields with masses $m_1, m_2$ and self-coupling constants $\lambda_1, \lambda_2$
accordingly, $V_0$ is a normalization constant. This potential was obtained in Ref. \cite{Dzhunushaliev:2006di}
at approximate modeling of condensate of a
gauge field in the SU(3) Yang-Mills theory. The potential (\ref{pot2}) has two global
minima at $\phi=0, \chi = \pm m_2$, local maximum at $\phi=0, \chi=0$  and two local minima at $\chi=0, \phi = \pm m_1$ at
values of the parameters $\lambda_1, \lambda_2$ used in this
paper. The conditions for existence of the
local minima are: $\lambda_1>0, m_1^2>\lambda_2 m_2^2/2$, and for the
global minima: $\lambda_2>0, m_2^2>\lambda_1 m_1^2/2$.

The inflationary models with several scalar fields are known (for example, the hybrid inflation model~\cite{Linde1},
see also the review~\cite{Lyth}).
The potential \eqref{pot2} was used at description of different models: in Ref.~\cite{Dzhun} for a model
of boson star, in Ref.~\cite{Dzhun1} at consideration of the cosmic string, in Ref.~\cite{Dzhun2}
at modeling of the phantom energy in the early Universe, in Ref.~\cite{Dzhun3} for a model of thick brane.

In this Letter we consider inflationary and scalaron stages within the framework
of the flat anisotropic model of the early Universe with the potential
\eqref{pot2}.

\section{General equations}
We start from the following Lagrangian:
\begin{equation}
\label{lagr}
L=-\frac{R}{16\pi G}+\frac{1}{2}\partial_{\mu}\phi \partial^{\mu}\phi+\frac{1}{2}\partial_{\mu}\chi \partial^{\mu}\chi-V(\phi, \chi).
\end{equation}
The anisotropic metric is choosing in the form:
\begin{equation}
\label{metr}
ds^{2}=dt^{2}-a_1^{2}(t)dx^{2}-a_2^{2}(t)dy^{2}-a_3^{2}(t)dz^{2},
\end{equation}
where $a_1,a_2,a_3$ are the scale factors on directions $x$, $y$,
and $z$ accordingly. For carrying out of calculations, it is
convenient to represent the scale factors in the following form~\cite{Gur1} (see also~\cite{Gur2}):
\begin{equation}
\label{scale}
a_1(t)=r(t)q_{1}, \quad a_2(t)=r(t)q_{2}, \quad a_3(t)=r(t)q_{3},
\end{equation}
where $q_{1},q_{2},q_{3}$ are dimensionless variables satisfying the
requirements:
\begin{equation}
\label{req}
\prod\limits_{\alpha =1}^{3}q_{\alpha }=1, \qquad \sum\limits_{\alpha
=1}^{3}\left( \dot {q}_{\alpha }/q_{\alpha }\right) =0,
\qquad \dot {q}_{\alpha}=\frac{d q_{\alpha}}{dt},
\end{equation}
whence it follows that $\prod\limits_{\alpha =1}^{3}a_{\alpha }=r^{3}$. For the line element \eqref{metr},
with  account of \eqref{scale}, the components of the
Ricci tensor will be:
\begin{eqnarray}
\label{Ricci}
-R_{0}^{0}&=&3\frac{\ddot {r}}{r}+\sum\limits_{\alpha =1}^{3}\left(
\frac{\dot {q}_{\alpha }}{q_{\alpha }}\right) ^{2},  \nonumber \\
-R_{\alpha }^{\alpha } &=&\frac{\ddot {r}}{r}+2\left( \frac{\stackrel{%
.}{r}}{r}\right) ^{2}+3\frac{\dot r}{r}\frac{\dot {q}%
_{\alpha }}{q_{\alpha }}+\left( \frac{\dot {q}_{\alpha }}{q_{\alpha }}%
\right) ^{\mbox{\bf .}}, \\
-R &=&6\left[ \frac{\ddot {r}}{r}+\left( \frac{\dot {r}}{r}%
\right) ^{2}\right] +\sum\limits_{\alpha =1}^{3}\left( \frac{\dot {q}%
_{\alpha }}{q_{\alpha }}\right) ^{2}.  \nonumber
\end{eqnarray}

Let us use the last expressions for obtaining of (1-1) and (2-2) components of the
Einstein tensor:
\begin{eqnarray*}
G_{1}^{1} &=&2\frac{\ddot {r}}{r}+\left( \frac{\dot {r}}{r}%
\right) ^{2}-3\frac{\dot {r}}{r}\frac{\dot {q}_{1}}{q_{1}}%
-\left( \frac{\dot {q}_{1}}{q_{1}}\right) ^{\mbox{\bf .}}+\frac{1}{2}%
\sum\limits_{\alpha =1}^{3}\left( \frac{\dot {q}_{\alpha }}{q_{\alpha
}}\right) ^{2}, \\
G_{2}^{2} &=&2\frac{\ddot {r}}{r}+\left( \frac{\dot {r}}{r}%
\right) ^{2}-3\frac{\dot {r}}{r}\frac{\dot {q}_{2}}{q_{2}}%
-\left( \frac{\dot {q}_{2}}{q_{2}}\right) ^{\mbox{\bf .}}+\frac{1}{2}%
\sum\limits_{\alpha =1}^{3}\left( \frac{\dot {q}_{\alpha }}{q_{\alpha
}}\right) ^{2}.
\end{eqnarray*}
Subtracting the component $G_{2}^{2}$ from $G_{1}^{1}$, one obtains:
\[
3\frac{\dot {r}}{r}\left( \frac{\dot {q}_{2}}{q_{2}}-\frac{%
\dot {q}_{1}}{q_{1}}\right) +\left( \frac{\dot {q}_{2}}{q_{2}}-%
\frac{\dot {q}_{1}}{q_{1}}\right) ^{\mbox{\bf .}}=0.
\]
Introducing in the last equation a notification $Q_{\alpha \beta }=\left(
\dot {q}_{2}/q_{2}-\dot {q}_{1}/q_{1}\right) $, we have:
\[
3\frac{\dot {r}}{r}+\frac{\dot {Q}_{\alpha \beta }}{Q_{\alpha
\beta }}=0,
\]
that gives after integration:
\[
Q_{\alpha \beta }=C_{\alpha \beta }/r^{3},
\]
where $C_{\alpha \beta }$ are integration constants. Hence it appears:
\begin{equation}
\label{eq5}
\frac{\dot {q}_{\alpha }}{q_{\alpha }}=\frac{C_{\alpha }}{r^{3}},
\end{equation}
where $C_{\alpha }$ are integration constants
and, according to the requirements (\ref{req}), $\sum\limits_{\alpha =1}^{3}C_{\alpha }=0$.
Integrating the last equation, one can obtain:
\begin{equation}
\label{eq6}
q_{\alpha }=A_{\alpha }\exp \left\{ C_{\alpha }\int \frac{dt}{r^{3}}\right\}
,
\end{equation}
where $A_{\alpha }$ are integration constants and $\prod\limits_{\alpha
=1}^{3}A_{\alpha }=1$. Now, using the relation (\ref{eq5}), we have from (\ref{Ricci}):
\begin{equation}
\label{eq7}
-R_{0}^{0}=3\frac{\ddot {r}}{r}+\frac{1}{r^{6}}\sum\limits_{\alpha
=1}^{3}C_{\alpha }^{2},\,\,\,\,\,-R=6\left[ \frac{\ddot {r}}{r}%
+\left( \frac{\dot {r}}{r}\right) ^{2}\right] +\frac{1}{r^{6}}%
\sum\limits_{\alpha =1}^{3}C_{\alpha }^{2},
\end{equation}
where the coefficient $\sum\limits_{\alpha =1}^{3}C_{\alpha }^{2}$ determines an anisotropy of the
model. Then one can obtain from Eqs. (\ref{eq7}) and \eqref{lagr} the (0-0) component of the Einstein's equations
(here and further we use the geometric units $8\pi G=1, c=1$):
\begin{equation}
\label{eq8}
3\left( \frac{\dot {r}}{r}\right) ^{2}-\frac{1}{2r^{6}}%
\sum\limits_{\alpha =1}^{3}C_{\alpha }^{2}=\frac{1}{2}\left[\dot\phi^2+\dot\chi^2\right]+V(\phi,\chi).
\end{equation}

The corresponding equations for the scalar fields can be obtained from the Lagrangian \eqref{lagr}. They will be:
\begin{eqnarray}
\label{sfe_1}
\ddot \phi +3\frac{\dot r}{r}\dot \phi&=&-\phi\left[2\chi^2+\lambda_1 \left(\phi^2-m_1^2\right)\right],\\
\ddot \chi +3\frac{\dot r}{r}\dot \chi&=&-\chi\left[2\phi^2+\lambda_2 \left(\chi^2-m_2^2\right)\right].
\label{sfe_2}
\end{eqnarray}

\begin{figure}[h]
\begin{center}
\fbox{
  \includegraphics[height=7cm,width=9cm]{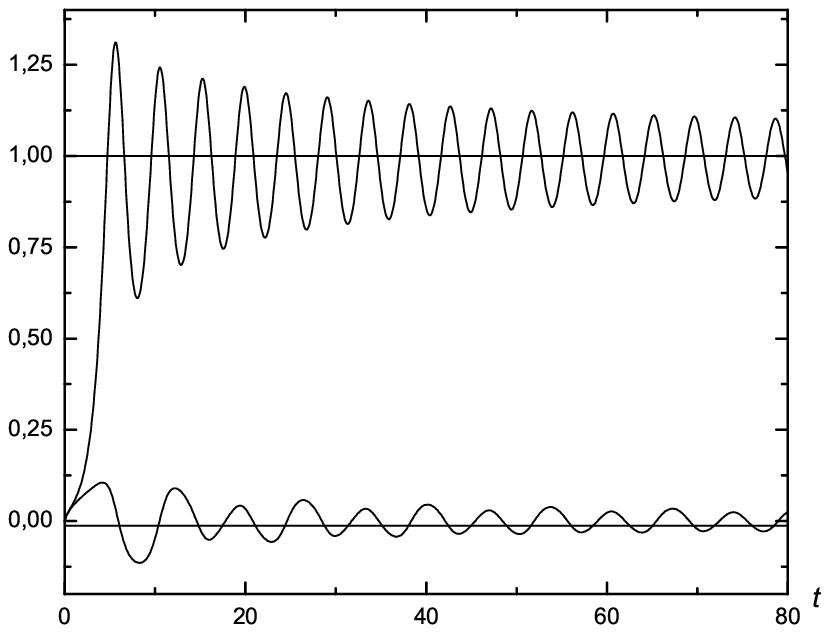}}
 \caption{The graphs for the scalar fields $\chi$ (top) and $\phi$ with $m_1=1, m_2=1, \lambda_1=0.1, \lambda_2=1$.
 Both fields tend asymptotically to the global minimum $\phi=0, \chi=m_2$.}
\label{exact_sol}
\end{center}
\end{figure}

 One can see that the system of equations \eqref{eq8}-\eqref{sfe_2} has solutions
beginning in the local maximum of the potential \eqref{pot2}, i.e. at $\phi=0, \chi=0$, and tending
asymptotically to the global minimum at $\phi=0, \chi = \pm m_2$ (Fig. \eqref{exact_sol}). (Here we choose the normalization constant
$V_0=(\lambda_1/4) m_1^4$ in order to provide $V(\phi, \chi)=0$ in the global minima.)
 These solutions are similar to
solutions in systems with spontaneous symmetry breaking~\cite{Linde}. In such systems, scalar fields roll down from maximum of
a potential (inflationary stage) toward minimum with subsequent fast oscillations (scalaron stage).

In the next section we will consider behavior of the model on these two stages.

\section{Inflationary and scalaron stages}
\subsection{Inflationary stage}

As is well-known~\cite{Linde}, the conditions
$\dot {\phi }^{2}, \dot {\chi }^{2}\ll V(\phi, \chi),\,\ddot {\phi}, \ddot {\chi }\ll
\partial V(\phi,\chi)/\partial (\phi, \chi)$ should be fulfilled on the stage of inflation. It allows to
rewrite Eqs. \eqref{eq8}-\eqref{sfe_2} as follows:
\begin{eqnarray}
\label{infl}
3\left( \frac{\dot {r}}{r}\right) ^{2}-\frac{1}{2r^{6}}\sum\limits_{\alpha =1}^{3}C_{\alpha }^{2}&=&V(\phi,\chi), \nonumber \\
3\frac{\dot r}{r}\dot \phi&=&-\phi\left[2\chi^2+\lambda_1 \left(\phi^2-m_1^2\right)\right],\\
3\frac{\dot r}{r}\dot \chi&=&-\chi\left[2\phi^2+\lambda_2 \left(\chi^2-m_2^2\right)\right]. \nonumber
\end{eqnarray}
We will search for inflationary solutions near the maximum of the potential \eqref{pot2} in
the following form:
\begin{equation}
\label{pert}
\phi=\delta \phi, \qquad \chi=\delta \chi, \qquad \delta \phi, \delta \chi \ll 1.
\end{equation}
The first equation from \eqref{infl} will then be:
\begin{equation}
\label{Ein00}
\left( \frac{\dot {r}}{r}\right) ^{2}-\frac{\sigma}{r^{6}}=H^2, \quad
\sigma=\frac{1}{6}\sum\limits_{\alpha =1}^{3}C_{\alpha }^{2}, \quad H^2=\frac{\lambda_2 m_2^4}{12},
\end{equation}
where $\sigma$ is an anisotropy parameter and $H$ is an effective Hubble constant. The last equation has a
solution:
\begin{equation}
\label{sol_Ein00}
r^{3}=\frac{\sqrt{\sigma }}{H}\sinh{3H(t-t_{0})},
\end{equation}
where $t_{0}$ is an integration constant.

Let us show that for isotropic case, i.e. at $\sigma \rightarrow 0$, Eq. \eqref{sol_Ein00} describes
usual exponential expansion. Indeed, if we take the arbitrary constant $t_{0}$ in the form
$t_{0}=\ln{\left( \sqrt{\sigma }H^{2}/2\right)}/3H $, then:
\begin{equation}
\label{isotr}
r^{3}=H^{-3}e^{3H t},
\end{equation}
i.e. one has usual inflationary solution for a flat isotropic model.

Let us further find  evolution of the scalar fields. Using \eqref{pert}, we can rewrite the second and
the third equations from \eqref{infl} as follows:
$$
3\frac{\dot r}{r}\delta \dot \phi-\lambda_1 m_1^2 \delta \phi=0, \quad
3\frac{\dot r}{r}\delta \dot \chi-\lambda_2 m_2^2 \delta \chi=0,
$$
or, taken into account \eqref{sol_Ein00}, we have:
\begin{eqnarray}
\delta \dot \phi-\frac{1}{3} \frac{\lambda_1 m_1^2}{H \coth{3H(t-t_0)}} \delta \phi&=&0, \\
\delta \dot \chi-\frac{1}{3} \frac{\lambda_2 m_2^2}{H \coth{3H(t-t_0)}} \delta \chi&=&0,
\end{eqnarray}
with solutions:
\begin{eqnarray}
\delta \phi&=& B_1 \left[\cosh{3H(t-t_0)}\right]^{\lambda_1 m_1^2/9H^2}, \\
\delta \chi&=& B_2 \left[\cosh{3H(t-t_0)}\right]^{\lambda_2 m_2^2/9H^2},
\end{eqnarray}
where a sign of the integration constants $B_1, B_2$ depends on whether the solution should appear in the
global minimum at $\phi=0, \chi=-m_2$ or at $\phi=0, \chi=+m_2$.

Now we can find a form of the metric coefficients $a_{\alpha }$. Eqs. \eqref{eq6} and \eqref{sol_Ein00} yield:
\begin{equation}
\label{eq18}
q_{\alpha }=A_{\alpha }\left[ \frac{e^{3H(t-t_{0})}-1}{
e^{3H(t-t_{0})}+1}\right]^{C_{\alpha }/3\sqrt{\sigma }},
\end{equation}
and, using \eqref{scale}, \eqref{sol_Ein00} and \eqref{eq18}, we finally have:
\begin{equation}
\label{eq19}
a_{\alpha }=A_{\alpha }\left[ \frac{e^{3H(t-t_{0})}-1}{
e^{3H(t-t_{0})}+1}\right]^{C_{\alpha }/3\sqrt{\sigma }}\left[ \frac{
\sqrt{\sigma }}{H}\sinh{3H(t-t_{0})}\right] ^{1/3}.
\end{equation}

\subsection{Scalaron stage}
On this stage, there are
fast oscillations of the fields near the minimum of the potential \eqref{pot2} with loss of the energy for
creation of particles (hot plasma). We will search for a solution of Eqs. \eqref{eq8}-\eqref{sfe_2} in the
following form:
\begin{equation}
\label{pert1}
\phi=\delta \phi, \quad \chi=m_2+\delta \chi,
\end{equation}
keeping in the potential terms up to second order on $\delta \phi, \delta \chi$. Then
Eqs. \eqref{eq8}-\eqref{sfe_2} and \eqref{pot2} yield:
\begin{eqnarray}
\label{sfe_3}
\delta \ddot \phi+3\frac{\dot r}{r} \delta \dot \phi+2\left[m_2^2-\frac{\lambda_1}{2}m_1^2\right]\delta \phi&=&0,\\
\label{sfe_4}
\delta \ddot \chi+3\frac{\dot r}{r} \delta \dot \chi+2\lambda_2 m_2^2\delta \chi&=&0,\\
\label{Ein_scal}
3\left( \frac{\dot {r}}{r}\right) ^{2}-\frac{1}{2r^{6}}
\sum\limits_{\alpha =1}^{3}C_{\alpha }^{2}&=&\frac{1}{2}\left[\delta \dot\phi^2+\delta \dot\chi^2\right]+
\left[m_2^2-\frac{\lambda_1}{2}m_1^2\right]\delta \phi^2+\lambda_2 m_2^2\delta \chi^2.
\end{eqnarray}
(Note that the expression $(m_2^2-(\lambda_1/2) m_1^2)>0$ due to the condition for existence of the global minima, see the Introduction.)

Let us search for solutions of Eqs. \eqref{sfe_3}-\eqref{sfe_4} by the following method~\cite{Gur2}:
\begin{equation}
\label{bog}
\delta \phi =A(t) \sin{\sqrt{2(m_2^2-(\lambda_1/2) m_1^2)}(t-t_0)}, \quad
\delta \chi =B(t) \sin{\sqrt{2\lambda_2 m_2^2}(t-t_0)},
\end{equation}
where $A(t), B(t)$ are some arbitrary functions obeying the following conditions on the scalaron stage:
$$\dot A/A \ll \sqrt{2(m_2^2-(\lambda_1/2) m_1^2)}, \quad \dot B/B \ll \sqrt{2\lambda_2 m_2^2}.$$ Then, inserting
\eqref{bog} into Eqs. \eqref{sfe_3}-\eqref{sfe_4} and neglecting terms with second derivatives, one
can obtain:
\begin{equation}
\label{AB}
A=D_1 r^{-3/2}, \quad B=D_2 r^{-3/2},
\end{equation}
where $D_1, D_2$ are integration constants. Now, inserting \eqref{bog} into Eq. \eqref{Ein_scal} and keeping only
zero terms on derivative, we have:
\begin{equation}
\label{Ein_scal_1}
3\left( \frac{\dot {r}}{r}\right) ^{2}-\frac{1}{2r^{6}}
\sum\limits_{\alpha =1}^{3}C_{\alpha }^{2}=
\left[m_2^2-\frac{\lambda_1}{2}m_1^2\right] A^2+\lambda_2 m_2^2 B^2,
\end{equation}
or, taken into account \eqref{AB},
\begin{equation}
\label{Ein_scal_2}
\dot r^2-\sigma r^{-4}-(\gamma_1+\gamma_2) r^{-1}=0,
\end{equation}
where
$$
\gamma_1=\frac{1}{3}D_1^2 \left[m_2^2-\frac{\lambda_1}{2}m_1^2\right], \quad
\gamma_2=\frac{1}{3}D_2^2 \lambda_2 m_2^2.
$$
Eq. \eqref{Ein_scal_2} has a solution:
\begin{equation}
\label{sol_r}
r^{3}=\left( \frac{9}{4}(\gamma_1+\gamma_2)^{2}(t-t_{0})^{2}-\sigma \right)/
\left(\gamma_1+\gamma_2 \right),
\end{equation}
where $t_0$ is an integration constant. Then $q_{\alpha }$ from \eqref{eq6} will take on form:
\begin{equation}
\label{eq25}
q_{\alpha }=A_{\alpha }\left[ \frac{3(\gamma_1+\gamma_2) (t-t_{0})/2-1}{3(\gamma_1+\gamma_2)
(t-t_{0})/2+1}\right]^{C_{\alpha }/3\sqrt{\sigma }},
\end{equation}
and, using \eqref{scale}, \eqref{sol_r} and \eqref{eq25}, one can obtain:
\begin{equation}
\label{eq26}
a_{\alpha }=A_{\alpha }\left[ \frac{3(\gamma_1+\gamma_2) (t-t_{0})/2-1}{3(\gamma_1+\gamma_2)
(t-t_{0})/2+1}\right]^{C_{\alpha }/3\sqrt{\sigma }}\left[ \left( \frac{9}{4
}(\gamma_1+\gamma_2)^{2}(t-t_{0})^{2}-\sigma \right) / (\gamma_1+\gamma_2) \right] ^{1/3}.
\end{equation}
Finally, from Eqs. \eqref{bog}, \eqref{AB} and \eqref{sol_r} we have:
\begin{eqnarray}
\label{eq27}
\delta \phi &=&D_1\left[ \left( \frac{9}{4}(\gamma_1+\gamma_2)^{2}(t-t_{0})^{2}-\sigma \right)
/(\gamma_1+\gamma_2) \right] ^{-1/2}\sin{\sqrt{2(m_2^2-(\lambda_1/2) m_1^2)}(t-t_0)},\\
\delta \chi &=&D_2\left[ \left( \frac{9}{4}(\gamma_1+\gamma_2)^{2}(t-t_{0})^{2}-\sigma \right)
/(\gamma_1+\gamma_2) \right] ^{-1/2}\sin{\sqrt{2\lambda_2 m_2^2}(t-t_0)}.
\end{eqnarray}

\section{Conclusions}
Summarizing, the anisotropic model of the early Universe on the inflationary and scalaron stages
was considered. It was shown that there are solutions beginning in the local maximum of the
potential \eqref{pot2} and tending asymptotically to one of the global minima. The solution near
the local maximum describes inflationary stage, and the solutions near the global minima - scalaron stage.
It is also obvious from \eqref{eq19} and \eqref{eq26} that the solutions  tend asymptotically to
isotropic case.

Note here that a type of the obtained solutions differs from the solutions from Refs.~\cite{Dzhun}-\cite{Dzhun3}.
In the case under consideration, we have exponential (near the local maximum) or oscillating solutions (near
the global minima) with arbitrary values of the parameters $m_1, m_2, \lambda_1, \lambda_2$. (They are restricted
only by the conditions on existence of the local and global minima.)
But in the referred above papers, the problems boiled down to finding of eigenvalues of the parameters $m_1, m_2$:
at other values of these parameters, the solutions are singular.


\begin{thebibliography}{99}
\bibitem{Linde}
A. Linde, {\it Particle Physics and Inflationary Cosmology}, Harwood, Chur (1990); hep-th/0503203.
\bibitem{Lidd}
A. R. Liddle and D. H. Lyth, {\it Cosmological inflation and large-scale structure}, Cambridge
University Press (2000).
\bibitem{Star}
A.A. Starobinsky, {\it Phys. Lett.} {\bf B91}, 99 (1980).
\bibitem{Dzhunushaliev:2006di}
  V.~Dzhunushaliev,
  ``Color defects in a gauge condensate'',
  hep-ph/0605070.
\bibitem{Linde1}
A. D. Linde, {\it Phys. Rev.} {\bf D49}, 748 (1994).
\bibitem{Lyth}
D. Lyth and A. Riotto, {\it Phys. Rept.} {\bf 314}, 1 (1999).
\bibitem{Dzhun}
V. Dzhunushaliev, ``Boson stars from a gauge condensate'', gr-qc/0604110.
\bibitem{Dzhun1}
V. Dzhunushaliev, V. Folomeev, K. Myrzakulov, R. Myrzakulov, {\it Mod. Phys. Lett.} {\bf A22}, 407 (2007);
gr-qc/0610111.
\bibitem{Dzhun2}
V. Dzhunushaliev, V. Folomeev, K. Myrzakulov, R. Myrzakulov,
``Bouncing off and inflation of the Universe with phantom fields'', gr-qc/0608025.
\bibitem{Dzhun3}
V. Dzhunushaliev,
``Thick brane solution in the presence of two interacting scalar fields'', gr-qc/0603020.
\bibitem{Gur1}
V.Ts. Gurovich and A. A. Starobinsky, {\it Sov. Phys. JETP}, {\bf 50}, 844 (1979).
\bibitem{Gur2}
V.N. Folomeev, V.Ts. Gurovich, {\it Gen. Rel. Grav.} {\bf 32}, 1255 (2000).
\end{thebibliography}
\end{document}